\documentstyle[12pt,fleqn,epsfig,amssymb]{article}
\newcommand {\bR}{{\Bbb R}}
\newcommand {\bN}{{\Bbb N}}
\newcommand {\bZ}{{\Bbb Z}}

\newcommand {\bT}{{\Bbb T}}

\newcommand {\cF}{{\cal F}}
\newcommand {\cG}{{\cal G}}
\newcommand {\cH}{{\cal H}}
\newcommand {\cI}{{\cal I}}
\newcommand {\cJ}{{\cal J}}
\newcommand {\cK}{{\cal K}}
\newcommand {\cL}{{\cal L}}
\newcommand {\cM}{{\cal M}} 
\newcommand {\cO}{{\cal O}} 
\newcommand {\cP}{{\cal P}} 
\newcommand {\cR}{{\cal R}} 

\newcommand{\beq}{\begin{equation}}
\newcommand{\eeq}{\end{equation}}
\newcommand{\Leq}[1]{\label{#1}\end{equation}}
\newcommand{\beqn}{\begin{eqnarray}}
\newcommand{\eeqn}{\end{eqnarray}}
\newcommand{\beqno}{\begin{eqnarray*}}
\newcommand{\eeqno}{\end{eqnarray*}}
\newtheorem{theorem}{Theorem} [section]

\newtheorem {remark}[theorem]{Remark}

\newtheorem {conject}[theorem]{Conjecture}

\renewcommand {\l}{\left}
\newcommand {\ri}{\right}
\newcommand {\r}{\right}
\newcommand {\LA}{\left\langle}
\newcommand {\RA}{\right\rangle}
\newcommand {\vep}{\varepsilon}
\newcommand {\vp}{\varphi}
\newcommand {\om}{\omega}
\newcommand {\ar}{\rightarrow}
\newcommand {\pa}{{\partial}}
\newcommand {\eh}{\textstyle \frac{1}{2}}
\newcommand {\ev}{\textstyle \frac{1}{4}}

\newcommand {\SE}{{\Sigma_E}}
\newcommand {\Vmax}{{V_{\rm max}}}
\newcommand {\Vmin}{{V_{\rm min}}}

\newcommand{\idty}{\hat{\rm 1\mskip-4mu l}} 
\newcommand{\OO}[1]{{{\cal O}\left(\hbar^{#1}\right)}}

\newcommand {\Eth}{{E_{\rm th}}}
\newcommand {\hP}{{\cP}}                    
\newcommand {\hH}{\hat{H}}  
\newcommand {\hI}{\hat{I}}
\newcommand {\hK}{\hat{K}}  

\newcommand {\hS}{\hat{S}}  
\newcommand {\hT}{\hat{T}}  

\newcommand {\hPhi}{\hat{\Phi}^t}

\newcommand {\hl}{\lambda}     
\newcommand {\bv}{{\bar{v}}}

\newcommand{\rstr}{{\!\hbox{
$\vert\mkern-4.8mu\hbox{\rm\`{}}\mkern-3mu$}}}
\newcommand {\qmbox}[1]{\quad\mbox{#1}\quad}
\newcommand {\Hhk}{{H^\hbar(k)}}                 
\newcommand {\Ehk}[1]{{E_{#1}^\hbar(k)}}         
\newcommand {\Enhk}{\Ehk{n}}                     
\newcommand {\phk}[1]{{\psi_{#1}^\hbar(k)}}      
\newcommand {\pnhk}{\phk{n}}                     

\newcommand {\lstar}{{\ell^\ast}}                
\newcommand {\mstar}{{m^\ast}}                
\newcommand {\Ehkq}[1]{{\tilde{E}_{#1}^\hbar(k)}}
\newcommand {\Enhkq}{\Ehkq{\lstar}}              
\newcommand {\pnhkq}{{\tilde{\psi}_\lstar^\hbar(k)}}
\newcommand {\phkq}[1]{{\tilde{\psi}_{#1}^\hbar(k)}}
\newcommand {\Lk}{{\Lambda^\hbar_I(k)}}          
\newcommand {\FLk}{{\cF\!\Lk}}                   
\newcommand {\GLk}{{\cG\!\Lk}}                   
\newcommand {\Pnk}{{P_\lstar^\hbar(k)}}
\newcommand {\vol}{{\rm vol}}
\newcommand {\dist}{{\rm dist}}
\newcommand {\NN}{\nonumber}
\newcommand {\po}{}
\newcommand {\tS}{\tilde{S}}
\begin{document}
\title {Quantum Transport on KAM Tori}
\author{Joachim Asch\thanks{CPT-CNRS, Luminy Case 907,
F-13288 Marseille Cedex 9, France. e-mail: asch@cpt.univ-mrs.fr}
\and
Andreas Knauf\thanks{Max-Planck-Institute for Mathematics in the
Sciences,
Inselstr.\ 22--26, D-04103 Leipzig, Germany.
e-mail: knauf@mis.mpg.de}
}
\date{November 1998}
\maketitle
\begin{abstract}
Although quantum tunneling between phase space tori occurs,
it is suppressed in the semiclassical limit $\hbar\searrow 0$
for the Schr\"{o}dinger equation of a particle in $\bR^d$
under the influence of a smooth periodic potential.

In particular this implies that the distribution of quantum group
velocities near energy $E$
converges to the distribution of the classical asymptotic velocities
near $E$, up to a term of the order $\cO(1/\sqrt{E})$.
\end{abstract}
\section{Introduction}
Consider abstractly a self-adjoint operator $H$ on its domain $D(H)$
in a Hilbert space $\cH$. Then for $\vep\geq 0$ one may call a pair
\[(\tilde{\psi},\tilde{E})\in D(H)\times \bR\qmbox{,}
\|\tilde{\psi}\|=1 \qmbox{,}
\|(H-\tilde{E})\tilde{\psi}\|\leq \vep\]
an $\vep$-{\em quasimode} \cite{Ar}.
In particular, eigenfunctions $\psi$ with eigenvalues $E$ are
$0$-quasimodes.

The existence of an $\vep$-quasimode $(\tilde{\psi},\tilde{E})$ implies
that the operator $H$ has spectrum $\sigma(H)$ in  $[\tilde{E}-\vep,
\tilde{E}+\vep]$.
In particular we are sure to find an eigenvalue $E$ in an interval
$[\tilde{E}-\mu,\tilde{E}+\mu]$ for $\vep\leq\mu$,
if we know that the spectrum in that interval is purely discrete.

If we know in addition that $E$ is the only such eigenvalue,
then, after
choosing an appropriate phase for its normalized eigenfunction $\psi$,
we have
\beq
\|\tilde{\psi}-\psi\|\leq \frac{2\vep}{\mu}.
\Leq{psi:near}

However, due to near-degeneracies of $\sigma(H)$ there may be no
eigenfunction of $H$ near $\tilde{\psi}$:\\[2mm]
The standard example is that of the Schr\"{o}dinger operator on the
line with the double well
potential $V(q):=(q-1)^2(q+1)^2$. Then for energies $E<1$ one may
construct
$\hbar^\infty$-quasimodes localized in one or the other well, whereas
all eigenfunctions of $H^\hbar$ exhibit parity.
Here the  near-degeneracy of the eigenenergies, which is of order
$\cO(\exp(-c/\hbar))$, is connected with tunneling between the two
components
of the energy shell of the classical system
(see, e.g.\ Lazutkin, \cite{La}).

So in that case phase space tunneling survives the semiclassical
limit, and one cannot confine a particle forever in a well.
As a physical consequence one may mention the $NH_3$ microwave
radiation.
\\[2mm]
For higher degrees of freedom $d$ these energy shell
components generalize to invariant
Lagrangian tori in phase space. If such invariant tori exist
and one has some control over the bicharacteristic flow in their
vicinity, then it is possible to
construct $\vep$-quasimodes of high accuracy ($\vep=\hbar^N$)
and thus to extract precise spectral informations in the semiclassical
limit
$\hbar\searrow0$, \cite{La}.

However, because of near-degeneracies in the spectrum,
in general one cannot draw any conclusion concerning the
semiclassical eigen{\em functions}.\\[2mm]
In recent years refined epitactic methods allowed to produce
semiconductors with periodic superlattices.
The electrons in these periodic potentials have a small effective
value of $\hbar$, leading to interesting effects (see \cite{WLR}).
In this context it is important to know to which extent one may model
the electronic behavior classically.

This motivates our study of Schr\"{o}dinger operators
\[H^\hbar = -{\hbar^2\over2}\Delta+V\qmbox{on}\cH:=L^2(\bR^d) \]
whose potential $V\in C^\infty(\bR^d,\bR)$ is periodic w.r.t.\ a
regular lattice $\cL\subset\bR^d$.

We may consider $V$ as a function $V:\bT\ar\bR$ on the
$d$-torus $\bT := \bR^d/ \cL$.

The $\cL$-invariance and the Bloch theorem imply that $H^\hbar$
conjugates unitarily to the direct integral of the operators
\beq
\Hhk := \eh (D+ \hbar k)^2 + V\qmbox{on}L^2(\bT)\qquad (k\in\bT^*),
\Leq{Hk}
acting on
$$\int^\oplus_{\bT^*}L^2(\bT,{dq}) \frac{dk}{\vol\bT^*},$$
where $\cL^\ast$ is the dual lattice with {\em Brillouin zone}
$\bT^{*}:=\bR^d/\cL^\ast$ and $D:=-i\hbar\nabla$ is the momentum
operator.
It follows that the spectrum
consists of bands.
Up to measure zero sets due to degeneracies, the eigenvalues $\Enhk$
are
analytic in $k$, and are non-constant,
(see, e.g.,Thomas \cite{thom}, Wilcox \cite{Wilc}, and Reed and Simon
\cite{RSiv}).
Thus the {\em group velocity} $\hbar^{-1}\nabla_k \Enhk$ vanishes
at most on a set of measure zero.

On the other hand the symmetry $E^\hbar_n(-k)=\Enhk$
of the band functions
implies in the non-degenerate case that the group velocity vanishes
for $k=0$ and the other
$2^d-1$ fixed points of $k\mapsto -k$ on $\bT^*$.

To see how this vanishing of the group velocity is connected with
phase space tunneling,
we consider the simplest case of
$d=1$ dimension (for $d=2$ see also \cite{DS}).

In that case the energy shell $\SE :=H^{-1}(E)\subset \cP$
of the Hamiltonian function
\[H(p,q):= \eh p^2+V(q)
\qmbox{on the phase space}\cP:=T^*\bT,\]
consists for energies
$E > \Vmax:=\max_q V(q)$ of two components, corresponding to ballistic
motion
to the right resp.\ to the left. These components are permuted by the
time
reversal transformation $(p,q)\mapsto (-p,q)$ on $\cP$.

As the eigenfunction $\pnhk$ can be chosen to be real for $k=0$,
it is semiclassically equally
concentrated on both (one-dimensional) tori corresponding to the energy
$E = E^\hbar_n(0)$.

The vanishing group velocity is one manifestation of that
fact. Thus for $k=0$, arbitrarily small values of $\hbar$ and
large times $t$ the quantum evolution
 $\exp(-i\Hhk t/\hbar)$ and
 the classical flow $\Phi^t:\cP\ar\cP$
generated by $H$ behave very differently.

However we argue that for general quasimomenta $k$ in $\bT^*$
{\em phase space tunneling is exceptional in the limit}
$\hbar\searrow0$.

More specifically, we conjectured in \cite{AK} that the quantum
distribution of
group velocities converges in the semiclassical limit to the
classical one, see Conjecture \ref{conjecture} below.

We proved this in \cite{AK} for the extreme cases of
potentials leading to ergodic motion, and for
separable potentials (which are the only known examples of periodic
potentials
leading to integrable motion).

Here we show a similar statement for arbitrary smooth potentials
and large
energies, where KAM tori are known to dominate the phase space
volume.\\[2mm]
After presenting the strategy in {\em Sect.\ 2},
we adapt in {\em Sect.\ 3\,} Lazutkin's results on KAM-quasimodes
to the present situation of a family $\Hhk$ of differential
operators. {\em Thm.\ \ref{thm:basic}\,}
contains our main result. It states that for
large energies $E$ a proportion $1-\cO(1/\sqrt{E})$ of the
eigenfunctions is semiclassically concentrated near a KAM torus.

This then leads to a corresponding statement
({\em Thm.\ \ref{thm:kam:v}\,}) for the semiclassical distribution of
group velocities, in accordance with the above conjecture.

In a final section, we try to abstract our strategy. We argue that
a mere {\em existence proof} for a full set of $\hbar^N$--quasimodes
with localized asymptotic velocities could imply the conjectured
classical limit of the distribution of group velocities.\\[5mm]
{\bf Acknowledgments.}
We thank Ruedi Seiler and SFB 288, TU Berlin, for
hospitality, for which  J.A.\ also thanks MPI in Leipzig.
\section{Heuristics}
Before we turn to formal statements and
proofs, we shortly describe the main ideas, starting with the following
observation.

Two given quasimodes associated to different KAM tori give rise to
different
expectations of the sub-principal symbol $\hbar k\cdot D$ of the
operator
$\Hhk $ defined in (\ref{Hk}). Thus they can be separated energetically
by varying the quasimomentum $k$, and for typical $k$ in the Brillouin
zone
$\bT^*$ one should not have too
many near-degeneracies of energies.

Of course we must consider scales in order to make this argument work.
In $d$ dimensions the mean spacing $E_{n+1}^\hbar(k)-\Enhk$ between
the eigenvalues of $\Hhk $ near $E>\Vmin$ is of the order $\hbar^d$.
Thus a priori one must consider in a fixed energy interval
about $\hbar^{-d}$ quasimodes
which may lead to a near-degeneracy with a given quasimode.
For $\hbar^N$-quasimodes we need an energy separation
of at least $\hbar^N$. So $N$ should be larger than $d$.

Such high precision
KAM quasimodes are constructed in the book \cite{La} by Lazutkin
(see also the article \cite{TW1}
by Thomas and Wassell for related results)
We apply this method after some straightforward adaptation to our
family (\ref{Hk}) of differential operators.

An important input for that construction consists in the refinement of
KAM theory presented in the paper \cite{Po} by P\"{o}schel.
Roughly speaking one uses that the deviation of
the Hamiltonian function $H$ from an
integrable one vanishes faster than any power of the phase space
distance to
the KAM set. In particular we may apply perturbative semiclassical
techniques in some $\hbar^\alpha$-neighborhood of the set of KAM tori.

A final remark concerns the phase space complement of the KAM set.
In general we do not have any information over individual eigenfunctions

and
eigenvalues concentrating semiclassically in that region.

In particular we cannot hope to lift
near-degeneracies between such eigenvalues and
the energies of the KAM-quasimodes by changing the quasi-momentum.
Moreover, if a quasimode is involved in such a near-degeneracy,
there need not be any eigenfunction $\pnhk$ near to that quasimode.

However, we can apply a box counting principle. We know  from KAM
theory that for large energies $E$ the complement of the
KAM set is of relative measure $\cO(1/\sqrt{E})$.

Then a Weyl argument implies that up to an exceptional set of relative
size
$\cO(1/\sqrt{E})$ the eigenvalues $\Enhk$
near $E$ are well-approximated by KAM quasimodes.

In the semiclassical limit these $\hbar^N$-quasimodes
$(\tilde{\psi},\tilde{E})$ are typically
energetically separated in the sense
that the associated intervals $[\tilde{E}-\hbar^N,\tilde{E}+\hbar^N]$
are disjoint. We have at least one eigenvalue $\Enhk$ in each such
interval.
Thus only an exceptional set of relative proportion $\cO(1/\sqrt{E})$
of these intervals may contain more than one
eigenvalue.

So for typical $k\in\bT^*$ most $\Enhk$ are not near-degenerate, and
thus the corresponding eigenfunctions $\pnhk$ are well approximated
by quasimodes $\tilde{\psi}$.
\section{KAM Estimates and Quasimodes}\label{kamestimates}
In order to apply KAM theory to $H$ with energies in
\beq
I:=[(1-\delta)E,(1+\delta)E]
\Leq{def:I}
near $E>0$, we change coordinates.
So consider the $d\times d$
matrix $L:=(\ell_1,\ldots,\ell_d)/(2\pi)$ of a basis
$(\ell_1,\ldots,\ell_d)$
for the configuration space lattice $\cL$, set $\hat{V}(\vp):=V(L\vp)$,
denote
by $\hat{\cP}:=T^*\hat{\bT}$ the phase space over the standard torus
\[\hat{\bT}:=\bR^d/(2\pi\bZ)^d\]
and define, using the matrix $M:= (L^tL)^{-1}$, the Hamiltonian
\[\hH_\vep:\hat{\cP}\ar\bR\qmbox{,}\hH_\vep(J,\vp):=\eh
(J,MJ)+\vep\hat{V}(\vp).\]
Then for the diffeomorphism
\[\cM_E: \cP \ar \hat{\cP}\qmbox{,} (p,q)\mapsto (J,\vp):=
\l( L^tp/\sqrt{E}, L^{-1}q \ri)\]
we have
\[E\cdot \hH_{1/E}\circ\cM_E=H,\]
and the flow $\hPhi_\vep$ generated by $\hH_\vep$
(w.r.t.\ the standard symplectic structure on
$\hat{\cP}$) is conjugate to the original flow, up to
a change of time scale:
\[\hat{\Phi}_{1/E}^{\sqrt{E}t}\circ \cM_E =
\cM_E\circ \Phi^t\qquad(t\in\bR).\]
$\hat{\Phi}_\vep^t$ becomes fully integrable
for perturbation parameter $\vep=0$. Namely
\[\hat{\Phi}_0^t(J_0,\vp_0)=(J_0,\vp_0+\om_0(J_0) t)\]
with the
{\em frequency vector}
\beq
\om_0 (J):=\frac{\pa \hH_0}{\pa J}.
\Leq{omo}
$\om_0$ is of independent variation, i.e.\ the matrix
\[\frac{\pa \om_0 (J)}{\pa J} = M\qmbox{is of rank}d.\]

So we are in a situation to apply KAM theory, see \cite{Po}.
For $\gamma>0$ and $\tau>d-1$ we consider the Diophantine sets
\beq
\Omega_\gamma:= \l\{\om \in\bR^d\mid \forall
k\in\bZ^d\setminus\{0\}:
|\om \cdot k|\geq \gamma \|k\|^{-\tau}\ri\}.
\Leq{Dio}
These are asymptotically of full measure as $\gamma\searrow 0$.

Denote the interval of new energies by
$\hI:=[1-\delta,1+\delta]$.
For $\vep=0$ the phase space region
$\hat{\cP}_\vep := \hH_\vep^{-1}(\hI)\subset \hat{\cP}$
is of the form
\[\hat{\cP}_0=\hat\cJ^\infty\times\hat{\bT}.\]
By KAM for $|\vep|$ small there exist a smooth generating function
$\hS_\vep$ on $\hat\cJ^\infty\times\hat{\bT}$ and a Hamiltonian
$\hK_\vep$
independent of the angle variables, with the following properties.
\begin{itemize}
\item
The frequency vector
\[\om_\vep:\hat\cJ^\infty\ar\bR^d\qmbox{,}
J^\infty\mapsto \nabla \hK_\vep(J^\infty)\]
is nondegenerate, and coincides for $\vep=0$ with (\ref{omo}).
\item
On the Cantor set
$\hat\cJ_{\gamma,\vep}^\infty:=(\om_\vep)^{-1}(\Omega_\gamma)$
of actions
\[\hH_\vep (J^\infty-\pa_\vp\hS_\vep(J^\infty,\vp),\vp)
= \hK_\vep(J^\infty)\qquad
\l((J^\infty,\vp)\in\hat\cJ_{\gamma,\vep}^\infty\times\hat{\bT}\ri).\]
\item
The symplectomorphism
\[\hT_\vep:\hat\cJ^\infty\times\hat{\bT}\ar \hat{\cP}\qmbox{,}
(J^\infty,\vp^\infty)\mapsto(J,\vp)\]
generated by $J^\infty\varphi-\hS_\vep(J^\infty,\vp)$ is near to
the identity.
\item
For $\gamma=c\sqrt{\vep}$ the set
$\hat{\cK}_\vep :=
\hT_\vep(\hat\cJ_{\gamma,\vep}^\infty\times\hat{\bT})
\cap\hat{\cP}_\vep$
of $\hat{\Phi}^t$-invariant KAM tori is of Liouville measure
\[\vol(\hat{\cK}_\vep)\geq
\vol(\hat{\cP}_\vep)\cdot \l(1-\cO(\sqrt{\vep}\ri)).\]
\item
The difference between the non-integrable Hamiltonian
function $\hH_\vep(x)$ and the integrable Hamiltonian
$\hK_\vep\circ\hT_\vep^{-1}(x)$ vanishes faster than any power
of the distance $\dist(x,\hat{\cK}_\vep)$ from the invariant tori,
and the same is true for any derivatives.
\end{itemize}
These statements imply corresponding results
about the symplectic map $T$ for the generating function
$S:=\sqrt{E}\hat{S}_{1/E}\circ\cM_E$ and the
integrable Hamiltonian $K:=E\cdot \hK_{1/E}\circ\cM_E$
\[\cJ_{\gamma,E}^\infty:=
\sqrt{E}(L^{t})^{-1}\hat\cJ_{\gamma,1/E}^\infty\]
and the subset
\beq\cK_I:= \cM_E^{-1}(\hat{\cK}_\vep )\subset \cP_I:=H_0^{-1}(I).
\Leq{kamset}
of KAM tori for the flow $\Phi^t$.
In particular,
\beq
\vol(\cK_I)\geq \vol(\cP_I)\cdot \l(1-\cO(E^{-1/2})\ri).
\Leq{vol:KI}

Turning to quantum mechanics,
the following theorem was essentially proven by Lazutkin in \cite{La}.
\begin{theorem}\label{quasimodes}
Let $\tau>2d$ in (\ref{Dio}), $0<\hbar<1$ and $k\in\bT^{\ast}$.
Define for $\alpha\in(1,{\tau-d\over d})$
$$\Lk:=\{\lstar\in \cL^\ast\mid
\dist(\hbar(\lstar+k),\cJ_{\gamma,E}^\infty)\le
\hbar^\alpha\}.$$
Then for $\beta:=1-\alpha d/(\tau-d)>0$
\begin{enumerate}
\item
\beq
(2\pi\hbar)^d\vert\Lk\vert=\vert\cK_I\vert+\OO{\beta}.
\Leq{kam:weyl}
\end{enumerate}
Furthermore for $N\in\bN$, $\hbar$ small enough and $\lstar\in\Lk$
there exists a $\hbar^{N+1}$--quasimode
$(\Enhkq , \pnhkq)$. It follows that:
\begin{enumerate}
\item[2.]
there is an eigenvalue $\Ehk{}$ of $H^\hbar(k)$
with
$$\vert\Ehk{}-\Ehkq{\lstar}\vert\le \hbar^{N+1};$$
\item[3.]
for  the spectral projection $P$ on
$(\Ehk{}-\hbar^p, \Ehk{}+\hbar^p)$
it holds
$$\Vert (\idty-{ P})\pnhkq\Vert\le{\hbar^{N+1-p}}.$$
\item[4.]
Let $N>2d+2$ and $0<p<N+1-d$.
Then $\forall\vep>0\ \exists\alpha$ such that the dimension
${\bf N}$  of the space of all these quasimodes projected to the
spectral
subspace of
$\bigcup_{\lstar}(\Ehk{}-\hbar^p, \Ehk{}+\hbar^p)$ meets the estimate
\beq
(2\pi\hbar)^d{\bf N}=\vert{\cK}_I\vert+\OO{1-\vep}.
\Leq{bigN}
\end{enumerate}
\end{theorem}
{\bf Proof.}
This is essentially Theorem 41.10 in \cite{La}. We specialize some
formal
aspects to our case -- i.e.\ the configuration manifold
has no boundary and the
invariant Lagrangian tori are diffeomorphically projecting
to the configuration torus, so that we do
not need a Maslov operator.\\
{\bf (ad 1):} This is Lazutkin's Proposition 40.2.\\
{\bf (ad 2 and 3):}
Let $E$ be so large, that the KAM results hold true. The
Ansatz
for the quasimodes is:
\beq\Ehkq{}=\sum_{j=0}^{N+1} \hbar^jE_{j}(k),
\quad\phkq{}(q)=e^{{i\over\hbar}(S(k,q)-\hbar\langle
k,q\rangle)}\sum_{j=0}^N\hbar^jA_j(k,q)
\Leq{quasimodeansatz}
with
\beqn
\hspace{-5mm} A_j(k,.)\in C^\infty(\bT)& ,&\NN \\
 \label{boundaryconditions}
\hspace{-5mm} S(k,.)\in C^\infty(\bR^d)&, &
S(k,q+\ell)-S(k,q)-\hbar\langle k,\ell\rangle\in 2\pi\bZ
\quad(\ell\in\cL).
\eeqn
Employing the operators
\beqno T_k:=-{i\over2}(\pa_qS\pa_q+\pa_q\pa_qS)=
-i((\nabla_qS)\cdot\nabla_q+ \eh\Delta S)),
\eeqno
one computes
\beqn\label{ansatz}
\lefteqn{e^{-{i\over\hbar}(S-\hbar\langle k,q\rangle)}
\l(H^\hbar(k)-\sum_{j=0}^{N+1}\hbar^jE_j \ri)
\phkq{}(q)= }\NN \\
& &\l( \eh(\pa_qS)^2+V-E_0\ri)\sum_{j=0}^N\hbar^jA_j+\NN \\
& &\sum_{j=0}^N\hbar^{j+1}T_kA_j
-{\eh} \sum_{j=1}^{N+1}\hbar^{j+1}\Delta A_{j-1}-
\sum_{j=0}^{2N}\hbar^{j+1}\hspace{-2mm}
\sum_{l=\max(0,j-N)}^{\min(N,j)}\hspace{-2mm}E_{j+1-l}A_l
\eeqn
and is led to consider the equations

$$H(\pa_qS(k,q),q)-E_0(k)=\OO\infty\eqno{(SC)_{-1}}$$
and for $0\leq j\leq N$, with $A_{{-1}}:=0$
$$T_kA_j(k)-\eh\Delta A_{j-1}(k)+\sum_{l=0}^j
E_{j+1-l}(k)A_l(k)=\OO\infty\eqno{(SC)_{j}}$$
with the boundary conditions specified in (\ref{boundaryconditions}).

The first step is to find a solution of the Hamilton-Jacobi equation
$(SC)_{-1}$. By KAM we know that there exists $ K\in
C^\infty(\cJ^\infty), S_{\po}\in C^{\infty}(\cJ^\infty\times\bT)$ such
that not only
$$H(P-\pa_qS_{\po}(P,q), q)=K(P)\quad
\hbox{on }(\pa_PK)^{-1}(\Omega_\gamma)\times\bT$$
but
\beq
H(P-\pa_qS_{\po}(P,q),q) = K(P) + {{\cO}
\l(\dist(P,(\pa_PK)^{-1}(\Omega_\gamma))^\infty\ri)}
\Leq{HPS}
on $\cJ^\infty\times\bT$ with all derivatives. Now set
\beqn
\tS(P,q)&:=& Pq-S_{\po}(P,q).\label{SqP}\\
S(\lstar,k,q)&:=&\tS(\hbar(\lstar+k),q),\quad
E_0(\lstar, k):=K(\hbar(k+\lstar))\NN
\eeqn
then defines a solution of $(SC)_{-1}$.

Using the same strategy the transport equations $(SC)_j$ are now solved
in two
steps: first solve the corresponding equation indexed by $P$
approximatively
near a KAM torus, then replace $P$ by $\hbar(\lstar+k)$ for
$\lstar\in\Lk$
and exploit flatness of the functions.

\medskip
Let $E$ be so large that $\pa_{qP}^2\tS(P,q)$ is non-degenerate.
$\vert\det \pa_{qP}^2\tS(P,q)\vert\ dq$ is (the coordinate
representation of) an invariant measure on a KAM torus $P=const$.
So, with $T_P$ denoting
the transport operator with respect to $\tS(P,q)$:
$$(\pa_q \tS \pa_q + (\Delta \tS))\vert\det
\pa_{qP}^2\tS(P,q)\vert=0\Longleftrightarrow
T_P\underbrace{\sqrt{\vert\det
\pa_{qP}^2\tS(P,q)\vert}}_{=:A_0(P,q)}=0.$$
For arbitrary $P$ it follows that
$T_PA_0(P,q)=
\cO\left(\dist(P,(\pa_PK)^{-1}(\Omega_\gamma))^\infty\right)$
so
\beq
A_0(q,\lstar,k):=A_0(q,\hbar(\lstar+k)), \qquad E_1:=0
\Leq{E1}
satisfy $(SC)_0$ for $\lstar\in\Lk$.\\[2mm]
By (\ref{E1}) we may now suppose that
$A_0(P,q),E_1(P)\ldots A_j(P,q),E_{j+1}(P)$ meet
$$(T_PA_{j'} - \eh \Delta A_{{j'}-1}+
\sum_{l=0}^{{j'}-1} E_{{j'}+1-l}A_l)(P,q)=
\cO\left(\dist(P,(\pa_PK)^{-1}(\Omega_\gamma))^\infty\right).$$
Then the structure of the equation for $A_{j+1},E_{j+2}$
is
\beq\label{transportp}T_PA_{j+1}(P,q)=f(P,q)+E_{j+2}(P)A_0(P,q).\eeq
This is satisfied for
$P\in(\pa_PK)^{-1}(\Omega_\gamma)$ by
$$E_{j+2}(P):=-\int_\bT A_0^{-1}f(P,q(P,Q))\ dQ$$
$$A_{j+1}(P,q(P,Q)):=A_0(P,q(P,Q))
\sum_{0\neq \lstar\in\cL^\ast}{(A_0^{-1}f)\hat{}(\lstar,P)\over
\langle\pa_P K(P),\lstar\rangle}e^{i\langle Q,\lstar\rangle}.$$
Here $q(P,Q)$ is given by the canonical diffeomorphism
$T:(P,Q)\mapsto(p,q)$ generated by
$\tS(P,q)$, and $g\mapsto \hat{g}$ the Fourier-Transform
\[\hat g(\lstar,P):=\int_\bT
g(P,Q)e^{-i\langle Q,\lstar\rangle}dQ.\]

Indeed, equation (\ref{transportp}) is equivalent to
\beqno&&
(-i\pa_q\tS\pa_q(A_0^{-1}A_{j+1})=A_0^{-1}f+E_{j+2})(P,q)
\Longleftrightarrow\\
&&-i{d\over
dt}A_0^{-1}A_{j+1}\circ\Phi^t(\pa_q\tS(P,q),q)\rstr_{t=0}=
(A_0^{-1}f+E_{j+2})(P,q)
\eeqno
where $\Phi^t$ is the Hamiltonian flow of $H$.
But $A_0^{-1}A_{j+1}\circ\Phi^t\circ T^{-1}=A_0^{-1}A_{j+1}\circ
T^{-1}\circ\Psi^t$ where $\Psi^t(P,Q)=(P,Q+\pa_PKt)$ is the flow
generated by $K$. So equation
(\ref{transportp}) is met by the above defined objects which are well
defined
and smooth if $P$ labels a KAM torus and have a Whitney extension to
$\cJ^\infty\times\bT$.
So by the same argument as before
$$A_{j+1}(\lstar,k,q):=A_{j+1}(q,\hbar(\lstar+k)),\quad
E_{j+1}(\lstar,k):=E_{j+1}(\hbar(\lstar+k))$$
satisfy $(SC)_{j+1}$.

Define now with the functions so obtained the quasimode
$(\phkq{\lstar},\Ehkq{\lstar})$ by the formula (\ref{quasimodeansatz})
with $\phkq{\lstar}$ normalized and the sum running up to $N$; the sum
for $\Ehkq{\lstar}$ runs up to $N+1$. We then have
\[(H^\hbar(k)-\Ehkq{\lstar})\phkq{\lstar}=\OO{N+2}\]
so choosing $\hbar$ small enough we get the assertion.
Items 2 and 3 follow by general considerations about quasimodes.\\
{\bf (ad 4):}
To
deduce
(\ref{bigN}) one has to estimate
$\langle\phkq{\lstar},\phkq{\mstar}\rangle$,
which is Lazut\-kin's Proposition 41.9. \hfill $\Box$
\begin{remark}{\rm
By \cite{Po} it suffices to assume that the potential $V\in
C^l(\bT,\bR)$ for $l\in \bN$ large enough.
}\end{remark}
\section{Approximation of Eigenfunctions}
Let the $\hbar^{2N}$--quasimodes
$\{(\pnhkq,\Enhkq)\}_{\lstar\in\Lk}$ be given by
Thm.\ \ref{quasimodes} and denote by
\[\Pnk\qquad(\lstar\in\Lk)\]
the spectral projector for $\Hhk$ and the interval
$[\Enhkq-\hbar^N,\Enhkq+\hbar^N]$.
For each $\lstar\in\Lk$ there is a nearby eigenvalue
\beq
\Enhk\qmbox{ with }  |\Enhk-\Enhkq|\leq \hbar^{2N}.
\Leq{near:e}
So for $\hbar<\hbar_0$ we know in particular that
$\dim(\Pnk)\geq 1$.

But since the quasimode construction is only based on the KAM part of
phase
space, it does not suffice to know that the quasimode energies $\Enhkq$
are separated from each other to ensure that the eigenenergies are
isolated.
Thus we consider the subset
\beq
\FLk:=\{\lstar\in\GLk\mid \dim(\Pnk)=1\}\qquad (k\in\bT^*).
\Leq{FLk}
of the index set
\[\GLk:=\l\{ \lstar\in\Lk\mid
|\Enhkq-\Ehkq{\ell'}| > 2\hbar^N\mbox{ for }
\ell'\in\Lk\setminus_{\{\lstar\}} \ri\},\]
We obtain a map
\[\cI_k:\GLk\to\bN\]
by setting $\cI_k(\lstar):=n$ for some $n$ meeting (\ref{near:e}).
This map is one-to-one.

Its restriction
to $\FLk $
is uniquely defined, since for $\lstar\in\FLk$ \ $\Pnk$
is the one-dimensional projector for the eigenfunction
$\psi^\hbar_{\cI_k(\lstar)}(k)$ of $\Hhk$ whose eigenvalue
$E^\hbar_{\cI_k(\lstar)}(k)$
lies in $[\Enhkq- \hbar^N,\Enhkq+ \hbar^N]$.

The index set $\GLk$ of the separated quasimodes
may be very small. For example it is even empty for $k=0$
in $d=1$ dimensions, if $\hbar>0$ is small enough.
However, its {\em mean} cardinality
\[\LA | \cG\!\Lambda^\hbar_I | \RA :=\int_{\bT^*} |\GLk |
\frac{dk}{\vol\bT^*}\]
over the Brillouin zone turns out to be asymptotic to
\[\LA | \cG\!\Lambda^\hbar_I | \RA \sim (2\pi\hbar)^{-d}\vol(\cK_I),\]
with the KAM subset $\cK_I$ as defined in (\ref{kamset}). This is the
reason why indices in $\FLk$ are abundant on the average; it holds:
\begin{theorem} \label{thm:basic}
For $\lstar\in\FLk\, ,\, k\in\bT^*$
and a suitable choice of phase of the eigenfunction
$\psi^\hbar_{\cI_k(\lstar)}(k)$,
\beq
\|\phk{\cI_k(\lstar)}-\pnhkq\|\leq 2\hbar^N.
\Leq{X}
For $N>d+2$ there is a $\beta>0$ such that for
$I:=[(1-\delta)E,(1+\delta)E]$ with $E>\Eth$
\beq
\vol(\cK_I)-\vol(\cK_I^c)-\cO_E(\hbar^\beta)\leq
(2\pi\hbar)^d \LA | \cF\!\Lambda^\hbar_I | \RA
\leq \vol(\cK_I)+\cO_E(\hbar^\beta),
\Leq{Y}
with $\cK_I^c:=\cP_I\setminus \cK_I$. In particular
\beq
\l|\frac{(2\pi\hbar)^d \LA | \cF\!\Lambda^\hbar_I | \RA}
{\vol(\cP_I)} - 1 \ri|\leq
\sqrt{\frac{\Eth}{E}}+\cO_E(\hbar^\beta).
\Leq{Z}
\end{theorem}
\begin{remark}{\rm
The Liouville measure of the thickened energy shell is of
order
\beq
\vol(\cP_I) = c(\delta)\cdot E^{d/2}\cdot (1+\cO(1/E)).
\Leq{phasespace:vol}
}\end{remark}
{\bf Proof.}
Estimate (\ref{X}) follows from (\ref{psi:near}) and Def.\ (\ref{FLk}),
since the $(\pnhkq,\Enhkq)$ are $\hbar^{2N}$-quasimodes.

The upper bound in (\ref{Y}) follows from the Lazutkin result
(\ref{kam:weyl}) for $|\Lk|$.

We claim that
\beq
(2\pi\hbar)^d \LA | \cG\!\Lambda^\hbar_I | \RA\geq
\vol(\cK_I)-\cO(\hbar^\beta).
\Leq{claim}
By (\ref{kam:weyl})
this follows from an estimate of the form
\beq
(2\pi\hbar)^d
\LA | \Lambda^\hbar_I\setminus \cG\!\Lambda^\hbar_I | \RA
= \cO(\hbar^\beta).
\Leq{B}
But
\beq
\LA | \Lambda^\hbar_I\setminus \cG\!\Lambda^\hbar_I | \RA
\leq \int_{\bT^*} \sum_{\ell_1\neq\ell_2\in\Lambda^\hbar_I}
\chi\l(\Ehkq{\ell_1}-\Ehkq{\ell_2}\ri) \frac{dk}{\vol\bT^*},
\Leq{A}
where $\chi(x):=1$ for $|x|\leq 2\hbar^N$ and $0$ otherwise.

For $E$ large and $\hbar<\hbar_0$
\beqno
\l|\nabla \l(\Ehkq{\ell_1}-\Ehkq{\ell_2}\ri)\ri|&\geq &
\eh\l|\nabla \l(E_0(\ell_1, k)-E_0(\ell_2, k)\ri)\ri|\\
&\geq& \ev \hbar^2 |\ell_1-\ell_2|\geq {cte.}  \hbar^2
\eeqno
uniformly for all $k\in\bT^*$ and
$\ell_1\neq\ell_2\in\Lambda^\hbar_I$.
Thus by the implicit function theorem the set of quasimomenta
$k\in\bT^*$
leading to a degeneracy
\[\Ehkq{\ell_1}=\Ehkq{\ell_2}\]
of quasi-energies forms a hypersurface, and
\[\int_{\bT^*}\chi\l(\Ehkq{\ell_1}-\Ehkq{\ell_2}\ri)
\frac{dk}{\vol\bT^*} =\cO(\hbar^{N-2}).\]
Since $| \Lambda^\hbar_I|$ is of order $\cO(\hbar^{-d})$,
the r.h.s.\ of (\ref{A}) is thus of order $\cO(\hbar^{-2d+N-2})$.
So for $N>d+2+\beta$ estimate (\ref{B}) holds true, implying
(\ref{claim}).

We estimate the number
\beq
\LA | \cF\!\Lambda^\hbar_I | \RA = \LA | \cG\!\Lambda^\hbar_I | \RA -
\LA | \cG\!\Lambda^\hbar_I\setminus \cF \!\Lambda^\hbar_I | \RA
\Leq{FGGF}
from below by using (\ref{claim}) and the relation
\beq
\l|\GLk\setminus \FLk\ri|\leq
\l|\Xi^\hbar_I(k)\setminus \cI_k(\GLk)\ri|,\qquad(k\in\bT^*)
\Leq{GFX}
where
\[\Xi^\hbar_I(k):=\{ n\in\bN\mid \Enhk\in I\}\]
is the index set of all eigenvalues in the interval $I$.

Estimate (\ref{GFX}) follows by noting that its l.h.s.\ equals
the number of intervals
\[ [\Enhkq-\hbar^N,\Enhkq+\hbar^N]\qmbox{for} \ell\in\GLk\]
containing two or more eigenvalues $\Enhk$ (counted with multiplicity).
By definition of $\GLk$ these
intervals are disjoint, and we have
\[\Ehk{\cI_k(\lstar)}\in [\Enhkq-\hbar^N,\Enhkq+\hbar^N],\]
so that
further eigenvalues must be indexed by an integer belonging to the set
which
appears on the r.h.s.\ of (\ref{GFX}).

The Weyl estimate
\[(2\pi\hbar)^d \l|\Xi^\hbar_I(k)\ri| =
\vol(\cP_I)+\cO(\hbar)\qquad (k\in\bT^*)\]
is uniform in $k$, since the slope of the band functions is bounded
above by
\[ |\nabla_k\Enhk|\leq \hbar\sqrt{2(\Enhk-\Vmin)}\]
and thus of order $\hbar$ if $\Enhk\in I$ (see \cite{AK}, Corr.\ 2.4).

Thus the r.h.s.\ of (\ref{GFX}) is bounded above by
\[\l|\Xi^\hbar_I(k)\setminus \cI_k(\GLk)\ri|\leq
(2\pi\hbar)^{-d}\vol(\cP_I) - |\GLk| -\cO(\hbar^{1-d}).\]

Inserting that upper estimate for (\ref{GFX}) in (\ref{FGGF}) and using
(\ref{claim}) proves the lower bound in (\ref{Y}).

Finally, estimate (\ref{Z}) follows from (\ref{Y}) and the result
\[\frac{\vol(\cK_I^c)}{\vol(\cP_I)}=\cO \l(1/\sqrt{E}\ri),\]
see (\ref{vol:KI}).
\hfill$\Box$
\section{Asymptotic Velocity}
As a consequence of Birkhoff's Ergodic Theorem for $\hl$--almost all
$x_0\in\hP$
\[\bv^\pm(x_0) := \lim_{T\ar\pm\infty} \frac{1}{T} \int_0^T
p(t,x_0) dt \]
exist and are equal ($\hl$ denoting the Liouville measure on $\cP$).
In this case we set $\bv(x_0):=\bv^\pm(x_0)$, and otherwise
$\bv(x_0):=0$, thus defining the {\em asymptotic velocity}
\[\bv: \hP\ar\bR^d\]
which is a $\hl$--measurable phase space function.

We are particularly interested in the energy dependence of
asymptotic velocity and thus introduce
the  {\em energy-velocity map}
\beq
A := (H,\bar{v}): \hP \ar \bR^{d+1}.
\label{def:A}
\eeq
$A$ is $\hl$--measurable and generates an image measure
$\nu := \hl A^{-1}$ on $\bR^{d+1}$.

On the other hand (see \cite{AK}) for almost all $k\in\bT^*$
the operator of asymptotic velocity
\[\bar v^\hbar(k):=\lim_{T\to\infty}{1\over T}\int_0^T
e^{i\Hhk t}(D+\hbar k)e^{-i\Hhk t}\ dt.\]
exists and is given by
\[{\bar v}^\hbar(k)=\sum P_m^\hbar(k)(D+\hbar k)P_m^\hbar(k)=
\sum \hbar^{-1}\nabla_k E^\hbar_m(k)P_m^\hbar(k)\]
with the eigenprojections
$P^\hbar_m(k)$ of $\Hhk$.

The {\em quantum asymptotic velocities} are
defined by
\[ \bar{v}^\hbar_n (k) := \l\{ \begin{array}{cl}
\hbar^{-1}\nabla_k E^\hbar_n(k) & \mbox{, gradient exists} \\
0 & \mbox{, otherwise.} \end{array}
\r.\]
We equip the {\em semiclassical phase space} $\hP^\hbar := \bN\times
\bT^*$
with the {\em semiclassical measure}
$\hl^\hbar := (2\pi\hbar)^{d} \mu_1\times \mu_2$,
where $\mu_1$ denotes counting measure on $\bN$ and $\mu_2$
Haar measure on the Brillouin zone $\bT^*$.

In order to compare classical and quantum quantities,
we introduce the {\em energy-velocity map}
\[A^\hbar: \hP^\hbar \ar\bR^{d+1}\quad{\rm with }\quad
A^\hbar(n,k):=(E^\hbar_n(k),\bar{v}^\hbar_n(k))\]
and the image measure $\nu^\hbar := \hl^\hbar (A^\hbar)^{-1}$.\\[2mm]
{\bf Example:}
For $V\equiv 0$ (free motion) $\nu^\hbar=\nu$
independent of the value of $\hbar$.\\[2mm]
In \cite{AK} we stated the following conjecture, which we proved
for smooth $V$ leading to integrable resp.\ to ergodic motion (see also
\cite{Kn2} for ergodic motions generated by Coulombic periodic $V$):
\begin{conject} \label{conjecture}
For all $\cL$--periodic potentials $V\in C^\infty(\bR^d,\bR)$
\[w^\ast\!-\!\lim_{\hbar\searrow 0} \nu^\hbar = \nu\]
(which means
$\lim_{\hbar\searrow 0}\int_{\bR^{d+1}} f(x)d\nu^\hbar(x)=
\int_{\bR^{d+1}} f(x)d\nu(x)$
for continuous functions $f\in C^0_0(\bR^{d+1},\bR)$ of compact
support).
\end{conject}
\begin{remark}{\rm
One may also consider the stronger conjecture with continuous {\em
bounded}
test functions $f$, that is weak convergence in the language of
probability
theory.
}\end{remark}
Here we obtain a statement which verifies the conjecture in the high
energy limit.
To this aim we introduce the {\em ballistic scaling}
\[f_E(e,v):= E^{-d/2}f(e/E,v/\sqrt{E})\qquad (E>0)\]
of a test function $f\in C^0_0(\bR^{d+1},\bR)$, so that $f_1=f$.
We notice that for $V\equiv 0$ we have
$\nu(E,v) = C\cdot\delta(E-\eh v^2)$ so that
\[\int_{\bR^{d+1}} f_E(x)d\nu(x)\equiv \int_{\bR^{d+1}} f(x)d\nu(x)
\qquad (E>0).\]
The result is
\begin{theorem} \label{thm:kam:v}
For all $f\in C^0(\bR^{d+1},\bR)$ with compact support in
$\bR^+\times \bR^d$ we have
\beq
\limsup_{\hbar\searrow 0}\l| \int_{\bR^{d+1}} f_E(x)d\nu^\hbar(x)-
\int_{\bR^{d+1}} f_E(x)d\nu(x)\ri| =\cO(1/\sqrt{E}).
\Leq{high:E}
\end{theorem}
{\bf Proof.}
By our assumption on $f$ there is an interval $I$ of the form
(\ref{def:I}) so that $I\times \bR^d$
strictly contains the support of $f_E$.

The index set of eigenenergies in $I$ splits into the disjoint union
\[\Xi^\hbar_I(k) =\Xi^\hbar_1(k)\cup \Xi^\hbar_2(k)\qmbox{with}
\Xi^\hbar_1(k) := \cI_k(\FLk).\]
By (\ref{Z}), the volume estimate (\ref{phasespace:vol})
and injectivity of $\cI_k$
\[(2\pi\hbar)^d \LA |\Xi^\hbar_2 | \RA
=\cO(E^{(d-1)/2})+\cO_E(\hbar^\beta),\]
so that
\[(2\pi\hbar)^d \int \sum_{n\in \Xi^\hbar_2(k)}
f_E(\Enhk,\bar v_n^\hbar(k))dk= \cO(1/\sqrt{E})+\cO_E(\hbar^\beta).\]
This leads to a contribution of order $\cO(1/\sqrt{E})$ to
(\ref{high:E}),
so
that we need only estimate the contribution of $\Xi^\hbar_1$.
By (\ref{kam:weyl})
$$\lim_{\hbar\searrow0} (2\pi\hbar)^d \LA | \Lambda^\hbar_I\setminus
\Lambda^\hbar_1| \RA = 0
\qmbox{for}
\Lambda^\hbar_1(k):=
\{\lstar\in\Lk\mid \hbar(\lstar+k)\in\cJ_{\gamma,E}^\infty\}.$$
So it suffices to consider the contribution of the index set
\[\Xi^\hbar_{1,1}(k):= \cI_k(\FLk\cap\Lambda^\hbar_1(k))
\subset\Xi^\hbar_1(k) . \]
The result (\ref{high:E}) then follows from the estimate
\beq
\bar{v}^\hbar_n (k)={\pa_P  K}(\hbar(\lstar+k))+\cO(\hbar).
\Leq{e0}
for $\lstar\in\FLk\cap\Lambda^\hbar_1(k)$ and $n:=\cI_k(\lstar)$
and the identity
\[\bar{v}(x)={\pa_P  K}(P)\qquad
(P\in\cJ^\infty, x\in T(\{P\}\times\bT))\]
for the group velocity on the KAM tori which we both prove now.

\medskip
By definition (\ref{FLk}) of $\FLk$,
the eigenvalue $\Enhk$ is non-degenerate so that
\beq
\bar{v}^\hbar_n (k) = \LA\pnhk,\bar v^\hbar(k)\pnhk\RA
\Leq{e1}

For $\phi$ in the ($k$--invariant) domain of $\Hhk$ and $E\in\bR$ we
have the estimate
\[{1\over 2}\Vert{\bar v}^\hbar(k)\phi\Vert^2\le\Vert(\Hhk-E)\phi\Vert
\Vert\phi\Vert+\Vert V-E\Vert\Vert\phi\Vert^2\qquad
(k\in\bT^*).\]

It follows from Theorem (\ref{thm:basic}) that
\[\Vert{\bar v}^\hbar(k)(\phk{\cI_k(\lstar)}-\pnhkq)\Vert=\OO{N}\]
which implies for the expectation
\beq
\hspace{-1cm}
\bar{v}^\hbar_n (k) = \LA\phk{n},(D+\hbar k)\phk{n}\RA
= \LA\phkq{\lstar},(D+\hbar k)\phkq{\lstar}\RA+\OO{N}.
\Leq{e2}
By construction of the quasimodes
\beq
\LA\phkq{\lstar},(D+\hbar k)\phkq{\lstar}\RA =
\int_{\bT} \pa_q \tS(P,q)\ d\mu_{P}(q)+\OO{}
\Leq{e3}
for $P:=\hbar(\lstar+k), \tS$ as defined in (\ref{SqP}), and
\[d\mu_P(q):= \frac{\pa_{qP}^2\tS(P,q)\ dq}
{\int_{\bT}\pa_{q'P}^2\tS(P,q')\ dq' }.\]
Finally from the Hamilton--Jacobi equation, since the
classical flow is
ergodic on the invariant torus indexed by $P$, and since
$d\mu_P$ is the invariant measure in $q$ coordinates, it holds
\[\!\!\bar{v}(x)=\int_{\bT} \pa_q \tS(P,q)\ d\mu_{P}(q)={\pa_P  K}(P)\qquad
(P\in\cJ^\infty, x\in T(\{P\}\times\bT)).\]
Thus (\ref{e0}) follows from (\ref{e1}), (\ref{e2}) and (\ref{e3}).
\hfill$\Box$
\begin{remark}{\rm
Actually we have proven in addition to Theorem \ref{quasimodes} that
$\phkq{\lstar}$ lead to joint quasimodes of $H^\hbar(k), \bar
v^\hbar(k)$, namely:
\[\Vert (\bar
v^\hbar-\pa_PK(\hbar(\lstar+k)))\phkq{\lstar}\Vert=\OO{}
\qquad(k\in\bT^*,\lstar \in\FLk).\]
}\end{remark}
\section{Beyond KAM}
Theorem \ref{thm:kam:v} gives a partial answer to Conjecture
\ref{conjecture}, based on the KAM region $\cK_I\subset\cP_I$.
But what happens in the complement $\cK_I^c\,$?
There the classical dynamics is very complicated in general,
since one may encounter there further KAM tori (not predicted by the
estimates), Cantori, elliptic and hyperbolic periodic orbits,
large ergodic components etc.

With the exception of the elliptic orbits,
there is no direct generalization of the above KAM methods, and thus
it seems hopeless to control the wavefunctions semiclassically
supported in that region.
However, as the following example shows, other methods may work.\\[2mm]
{\bf Example.} Consider $d=2$ dimensions. As shown in \cite{AK},
in the presence of at least two geometrically distinct KAM tori
the motion on $\SE$ is ballistic ($\bv\neq 0$).
This is caused by the fact that these tori have codimension one in
$\SE$ and thus confine the flow between them.
We denote by $\cR_I\subset \cP_I$ the phase space region enclosed
by two nearby KAM tori (or rather families of such tori
indexed by the energy in $I$).

Using microlocal techniques, Shnirelman showed in \cite{Sh} the
existence
of a large number of quasimodes concentrated in $\cR_I$, see also
\cite{Co}.
Now for large energy $E$ the variation of $\bv$ w.r.t.\ the
restriction of Liouville measure to $\cR_I$ is small
in comparison with $E$.
Thus by Egorov's Theorem
the above quasimodes have group velocities near the classical $\bv$
values (see also \cite{AK}, Sect.~5).

Different such regions $\cR_I$, however, have different
classical asymptotic velocities. Thus
one should be able to apply the heuristics developed in Sect.~2
to that case, too ---
{\em without explicitly knowing the quasimodes}. \\[2mm]
When trying to work on this kind of arguments, one is led to
the paradoxical conclusion that sometimes it is more useful to
know quasimodes (with certain additional properties) of an operator
than to know its eigenfunctions.

To explain this,
consider the algebra generated by
\[\{H^\hbar(k), \bar v_1^\hbar(k),\ldots,\bar v_d^\hbar(k)\},\]
$\bar v_i^\hbar$ being the components of the operator of asymptotic
velocity -- which commute with $H^\hbar(k)$ --
and try to show the existence of
{\em joint} quasimodes. Arguing along the lines of Sect.\ 2, such an
existence
proof could suffice to prove Conjecture \ref{conjecture} in full
generality.
\end{document}